\documentclass[12pt,oneside]{article}
\usepackage{epsfig, cite}
\usepackage{color}
\usepackage{ifthen}
\usepackage{graphicx}

\def\p{\partial}

\def\half{{1\over 2}}
\def\({\left(}
\def\){\right)}
\def\[{\left[}
\def\]{\right]}

\def\te{\tilde{\eta}}

\def\e{\begin{equation}}
\def\q{\end{equation}}
\def\m{\begin{eqnarray}}
\def\n{\end{eqnarray}}

\begin{document}
\thispagestyle{empty} \setcounter{page}{0}
\renewcommand{\theequation}{\thesection.\arabic{equation}}

\vspace{2cm}

\begin{center}
{\Large{Slow-roll reconstruction for running spectral index}}

\vspace{1.4cm}

Qing-Guo Huang

\vspace{.2cm}

{\em School of physics, Korea Institute for
Advanced Study,} \\
{\em 207-43, Cheongryangri-Dong,
Dongdaemun-Gu, } \\
{\em Seoul 130-722, Korea}\\
\end{center}

\vspace{-.1cm}

\centerline{{\tt huangqg@kias.re.kr}} \vspace{1cm}
\centerline{ABSTRACT}
\begin{quote}
\vspace{.5cm}

We reconstruct from the WMAP three-year data the slow-roll
parameters in the standard slow-roll inflation model and the
noncommutative inflation model. We also investigate the evolution of
these slow-roll parameters. Requiring that slow-roll inflation lasts
more than 20 e-folds after CMB scales leave the horizon, we find
that the potential at the last stage of inflation takes the form
$V(\phi)=V_0(1+{\eta_c\over 2}{(\phi-\phi_c)^2\over M_p^2})$, where
$\eta_c$ is a constant. A natural mechanism to end inflation at
$\phi=\phi_c$ is the hybrid type inflation.

\end{quote}
\baselineskip18pt


\newpage

\setcounter{equation}{0}
\section{Introduction}

Inflationary models \cite{infl} not only explain the large-scale
homogeneity and isotropy of the universe, but also provide a natural
mechanism to generate the observed magnitude of inhomogeneity.
During the period of inflation, quantum fluctuations are generated
within the Hubble horizon, which then stretch outside the horizon to
become classical. In the subsequent deceleration phase after
inflation these frozen fluctuations re-enter the horizon, and seed
the matter and radiation density fluctuations observed in the
universe.

There are two notable problems in cosmology. One is how to extract
the amplitude of the power spectrum, including its scale dependence,
from observational data. The other is to understand the origin of
the inflationary models from the fundamental theory. In the last
decade hundreds of inflationary models have been proposed: see for
example \cite{ll}. However we still can not pick out a realistic
model among them. With the development of cosmological observations,
such as WMAP \cite{wmapt} and Sloan Digital Sky Survey \cite{sdss}
etc., we believe that now it is time to ask what the observations
imply for the above problems. A useful method to try to answer this
question is the slow-roll reconstruction, which has been
investigated for more than ten years \cite{rec,sll,chhk,hh}.
Unfortunately, till now no suggestive implications have been worked
out for the inflaton potential. In this paper we try to improve this
status.

Now the $\Lambda$CDM model is still an excellent fit to the WMAP
three-year data. A nearly scale-invariant, adiabatic primordial
power spectrum generated during inflation can be taken as the seeds
for the anisotropy of CMB. Even though a red power spectrum for the
curvature perturbations is a nice fit to the WMAP data, a running
spectral index slightly improves the fit \cite{wmapt}. The values of
the spectral index and its running and the upper bound on the
tensor-scalar ration are respectively \cite{wmapt} \e
n_s=1.21^{+0.13}_{-0.16},\quad
\alpha_s=-0.102^{+0.050}_{-0.043},\quad r\leq 1.5, \q at
$k=0.002$Mpc$^{-1}$. Combining WMAP with SDSS, a more stringent
constraint on the tensor perturbations $r\leq 0.67$ is obtained. The
running is allowed not only in the WMAP data, but also in
combination with other CMB and/or large scale structure information,
such as 2dFGRS \cite{tdfgrs} and the Sloan Digital Sky Survey
\cite{sdss}. Further analysis of a possible running spectral index
is discussed in \cite{fxy}. Many theoretical explorations of WMAP
results are carried out recently in
\cite{hls,mm,hld,hg,pr,eas,pes,frm}.

The WMAP data also implies that a red power spectrum ($n_s<1$) at
$k=0.05$Mpc$^{-1}$ is favored and the running of the spectral index
is not required at more than the $95\%$ confidence level. Thus not
only the spectral index runs, but the running is also
scale-dependent. In \cite{hg}, the author suggested a new parameter,
the running of running, which is defined as \e
\beta_s={d\alpha_s\over d\ln k}, \q to characterize the
scale-dependent running. The value of $\beta_s$ is estimated at the
linear approximation level to be \e \label{wtrr} \beta_s\simeq
{\Delta \alpha_s\over \Delta \ln k}\simeq{0-(-0.102)\over
\ln(0.05/0.002)}\simeq 0.0318. \q A running of running with the
order of magnitude $10^{-2}$ is expected. In \cite{pes}, the authors
also pointed out that the running of the spectral index should not
be a constant. In fact, nowadays data cannot provide the running of
running at a significantly statistical confidence level. If we
consider the constraint from the running of running, the
reconstruction is just a sub-set for the case without this
constraint.

We use the Horizon Flow Functions \cite{rec,sll,leach} to compute
the spectral index and its running and running of running. One of
the advantages of this approach is that it directly translates
between the slow-roll parameters and the running spectral index. A
further virtue of this approach is that when the slow-roll hierarchy
is truncated at finite order, the truncation is preserved by the
evolution equations.

In \cite{wmapt} the scale $0.002$ Mpc$^{-1}$ is used, which is close
to the observable horizon. However, this scale is not directly
related to a fundamental scale with the viewpoint of fundamental
theory. It is interesting for us to study the evolution of the
reconstruction at different scale. On the other hand, for the
slow-roll approximation, the location on the inflationary potential
corresponding to the observed perturbations is usually quantified by
the number of e-folds before the end of inflation as \cite{alb} \e
N(k)=-\ln{k\over a_0H_0}+{1\over 3}\ln{\rho_{reh}\over
\rho_{end}}+{1\over 4}\ln {\rho_{eq}\over \rho_{reh}}+\ln\sqrt{8\pi
V_k\over 3M_p^2}{1\over H_{eq}}+\ln 219\Omega_0h. \q A plausible
estimation in the second paper of \cite{alb} requires that the
slow-roll inflation lasts at least 20 e-folds after the observable
Universe leaves the horizon.

In this paper, we only fit the central value of the observational
parameters, which clearly implies that the the evolution of the
slow-roll parameters should be very important for the
reconstructions. In order to get more precise results, we should use
Monte Carlo Markov Chain analysis. We hope one can do it in the
future. Our paper is organized as follows. In section 2, we
calculate the spectral index and its running and running of running
in standard slow-roll inflationary models and reconstruct the
slow-roll parameters. In section 3, we compute the modification of
running spectral index in noncommutative inflationary models and
reconstruct the slow-roll parameters and the noncommutative
parameter. The evolution of the slow-roll parameters are also
investigated in section 2 and 3. In section 4, we suggest a
tentative potential for inflaton at the last stage of inflation.
Section 5 contains some concluding remarks.

\setcounter{equation}{0}
\section{Spectral index in usual inflation model}

In this paper, we only focus on the inflationary models governed by
single inflaton field $\phi$. In FRW universe, Friedmann equation
and the equation of motion for the homogeneous inflaton $\phi$ are
respectively \m H^2&=&{1\over 3M_p^2}\(\half {\dot
\phi}^2+V(\phi)\), \\ \ddot \phi&=&-3H\dot \phi-V', \n where $M_p$
is the reduced Planck mass scale, $V(\phi)$ is the potential of
inflaton. When ${\dot \phi}^2\ll V$ and $|\ddot \phi|\ll 3H|\dot
\phi|$, inflaton slowly rolls down its potential. We define for
convenience the slow-roll parameters as \m
\epsilon&\equiv&{3{\dot \phi}^2\over 2}\(V+\half {\dot \phi}^2\)^{-1}=2M_p^2\({H'\over H}\)^2, \\
\eta&\equiv&-{\ddot \phi\over H\dot \phi}=2M_p^2{H''\over
H},\label{hht} \n where prime denotes the derivative with respect to
$\phi$. The slow-roll conditions become \e \epsilon\ll 1, \quad
\hbox{and} \quad |\eta|\ll 3.\q In the whole paper, we assume,
without loss of generality, $\dot \phi<0$, so that $H'/H>0$. The
number of e-folds $N$ before the end of inflation is given by \e
\label{nph}{dN\over d\phi}=-{H\over \dot \phi}={1\over
\sqrt{2\epsilon}M_p}.\q The slow-roll parameter $\epsilon$
determines how fast the inflaton field evolves. If $\epsilon\ll 1$,
inflaton rolls down its potential very slowly.

Expand around an exact solution to first order, the amplitudes of
the scalar and tensor power spectra are respectively
\cite{rec,ll,lpb,lr,sl} \m \Delta_{\cal
R}^2&=&\(1-(2C+1)\epsilon+C\eta\)^2{H^2/M_p^2\over 8\pi^2\epsilon},
\\ \Delta_T^2&=&\(1-(C+1)\epsilon \)^2{H^2/M_p^2\over \pi^2/2}, \n
where $C=-2+\ln 2+\gamma\simeq -0.73$ and $\gamma$ is the Euler
constant originating in the expansion of the gamma function. The
tensor-scalar ratio is \e \label{tsr} r={\Delta_T^2\over
\Delta_{\cal R}^2}=16\epsilon(1+2C(\epsilon-\eta)). \q Taking into
account the slow-roll condition, we find \e \label{dkp}{d\over d\ln
k}=-M_p{\sqrt{2\epsilon}\over 1-\epsilon}{d\over d\phi}.\q The
evolution of the slow-roll parameters are given by \m
{d\epsilon\over d\ln k}&=&{2\epsilon^2-2\epsilon\eta\over
1-\epsilon}, \label{va}\\ {d\eta \over d\ln
k}&=&{-\xi+\epsilon\eta\over 1-\epsilon}, \label{vb}\\ {d\xi \over
d\ln k}&=&{2\epsilon\xi-\eta\xi-\zeta\over 1-\epsilon}, \label{vc}
\\ {d\zeta\over d\ln k}&=&{-\theta-2\eta\zeta+3\epsilon\zeta\over
1-\epsilon}, \label{vd}\n where \m
\xi&\equiv&4M_p^4{H'H^{(3)}\over H^2}, \\
\zeta&\equiv&8M_p^6{H'^2H^{(4)}\over H^3}, \\
\theta&\equiv&16M_p^8{H'^3H^{(5)}\over H^4},\n and $H^{(n)}=d^n
H/d\phi^n$. The above equations for the evolution of the slow-roll
parameters preserve the slow-roll hierarchy truncations. For
example, if $\xi=\zeta=\theta=...=0$, $d\xi/d\ln k=d\zeta/d\ln
k=...=0$.

Thus the spectral index and its running and running of running
respectively take the form \m n_s&\equiv&{d\ln \Delta_{\cal
R}^2\over d\ln
k}=1-4\epsilon+2\eta-2C\xi-8(C+1)\epsilon^2+(6+10C)\epsilon\eta,
\label{cns}
\\ \alpha_s&\equiv&{dn_s\over d\ln k}=-2\xi-8\epsilon^2+10\epsilon\eta-(8+14C)\epsilon\xi+2C\eta\xi+2C\zeta, \label{crun}\\
\beta_s&\equiv&{d\alpha_s\over d\ln
k}=-14\epsilon\xi+2\eta\xi-2C\xi^2+2\zeta-2C\theta\nonumber
\\
&-&32\epsilon^3+62\epsilon^2\eta-20\epsilon\eta^2-(46+56C)\epsilon^2\xi+(26+48C)\epsilon\eta\xi
\nonumber \\
&-&2C\eta^2\xi+(10+20C)\epsilon\zeta-6C\eta\zeta. \label{crrun} \n
We will use (\ref{tsr}), (\ref{cns}), (\ref{crun}) and (\ref{crrun})
to reconstruct the slow-roll parameters for the running spectral
index. In order to describe the evolution of the slow-roll
parameters, we introduce a new number of e-folds $n$ which is
related to the value of the scale factor by \e a(n)=a_{0.002}e^n, \q
where $a_{0.002}$ is the scale factor at the time when the comoving
fluctuation mode with wave number $k=0.002$Mpc$^{-1}$ crossed the
Hubble horizon during inflation. Using eq. (\ref{nph}), (\ref{dkp})
and $dn=-dN$, we find \e (1-\epsilon){d\over d\ln k}={d\over dn}. \q
For $k=0.05$Mpc$^{-1}$, $n_{0.05}\simeq \ln (0.05/0.002)=3.2$. In
the following, we will reconstruct the slow-roll parameters with
different slow-roll hierarchy truncations.

\subsection{$\xi=\zeta=\theta=0$}

In the first case we set $\xi=\zeta=\theta=0$. There are two free
parameters $\epsilon$ and $\eta$ to fit the two observed parameters
$n_s$ and $\alpha_s$. Solving eq. (\ref{cns}), we find \e
\eta={s+4\epsilon+8(1+C)\epsilon^2\over 2+(6+10C)\epsilon}, \q where
$s=n_s-1$. The running of the spectral index becomes \e
\alpha_s={\epsilon(5s+12\epsilon+16\epsilon^2)\over
1+(3+5C)\epsilon}, \q which is positive for the blue spectrum
$(n_s>1)$ if $\epsilon<-1/(3+5C)\simeq 1.5$. Slow-roll condition
requires $\epsilon\ll 1$ and the running of spectral index is
positive for blue tilt power spectrum. Thus two slow-roll parameters
$\epsilon$ and $\eta$ can not provide a negative running of the
spectral index with large absolute value.

\subsection{$\zeta=\theta=0$}

Here we take the thirst slow-roll parameter $\xi$ into account,
while $\zeta$ and $\theta$ are still set to zero. First we use three
slow-roll parameters $\epsilon$, $\eta$ and $\xi$ to fit the
spectral index and its running. Taking the spectral index and its
running as input, we show the constraint on these slow-roll
parameters and corresponding running of running in fig.1.
\begin{figure}[h]
\begin{center}
\epsfxsize=0.3\columnwidth \epsfbox{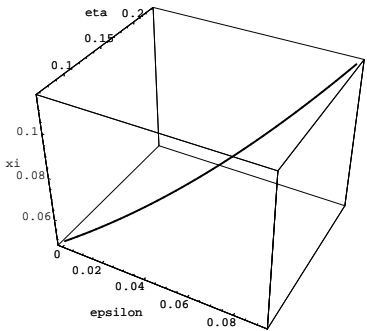}
\end{center}
\caption{Ignoring $\zeta$ and $\theta$, the constraints on the
slow-roll parameters and the corresponding running of running are
showed for $n_s=1.21$ and $\alpha_s=-0.102$, where $r\simeq
16\epsilon<1.5$.}
\end{figure}
They can be used to fit the spectral index and its running. However
they are roughly evaluated at the same order of magnitude. We do not
reach a reliable truncation of the slow-roll hierarchy and higher
order parameters should be taken into account.

The evolution of slow-roll parameters are governed by eq.
(\ref{va}), (\ref{vb}), (\ref{vc}) and (\ref{vd}). Solving these
equations, we figure out how the slow-roll parameters and spectral
index and its running evolve. Since there are only two constraints
($n_s$ and $\alpha_s$), but three free parameters, we can not fix
the value of $\epsilon$, $\eta$ and $\xi$ at $k=0.002$Mpc$^{-1}$
(or, $n=0$). We scan the corresponding value along the line in fig.
1. Here we pick out three points in fig. 1 and show the evolution of
the slow-roll parameters and the spectral index and its running
respectively in fig. 2.
\begin{figure}[h]
\begin{center}
\epsfxsize=0.6\columnwidth \epsfbox{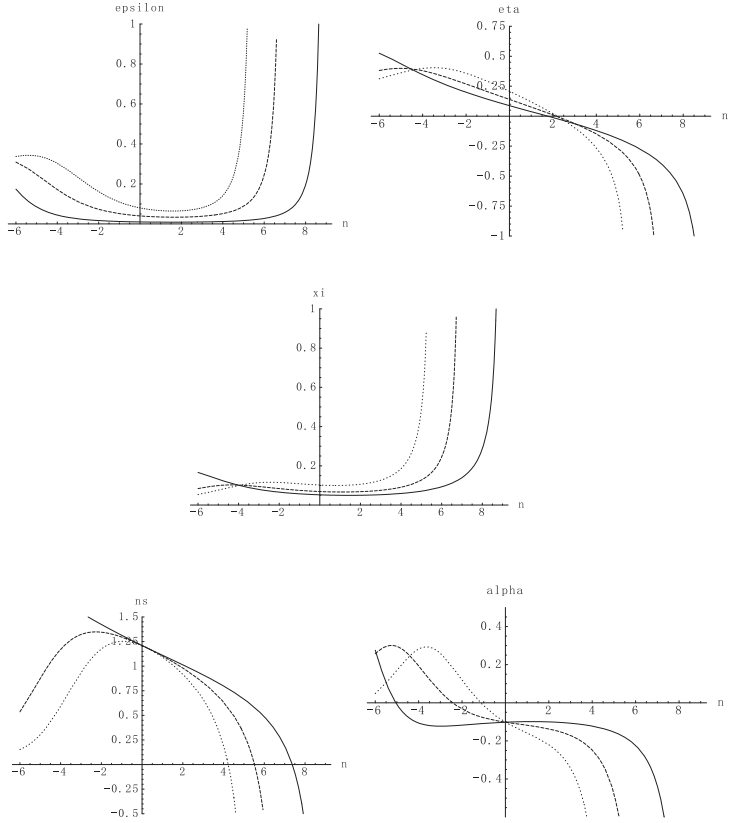}
\end{center}
\caption{Ignoring $\zeta$ and $\theta$, the evolution of the
slow-roll parameters and spectral index and its running are showed.
The solid lines correspond to $\epsilon=0.01, \eta=0.088, \xi=0.052$
at $k=0.002$Mpc$^{-1}$; the dashed lines correspond to
$\epsilon=0.04, \eta=0.140, \xi=0.069$; The dotted lines correspond
to $\epsilon=0.08, \eta=0.208, \xi=0.102$. }
\end{figure}
Similar behaviors for them are obtained for the other values of the
slow-roll parameters at $k=0.002$Mpc$^{-1}$. We find from fig. 2
that a large enough number of e-folds after the CMB scales leave the
horizon can not be achieved. In \cite{eas,pes}, the authors also
pointed out that the running of spectral index cannot be obtained if
we only consider these three slow-roll parameters and require that
the slow-roll inflation lasts not shorter than 30 e-folds after the
CMB scales leave the horizon.

\subsection{$\theta=0$}

In this case we take $\zeta$ into account. In \cite{chhk} the
authors also suggested this slow-roll parameter $\zeta$ for a
running spectral index. Since there are four free parameters, we
also consider the constraint from the running of running. Our
results are shown in fig. 3.
\begin{figure}[h]
\begin{center}
\leavevmode
\epsfxsize=0.3\columnwidth \epsfbox{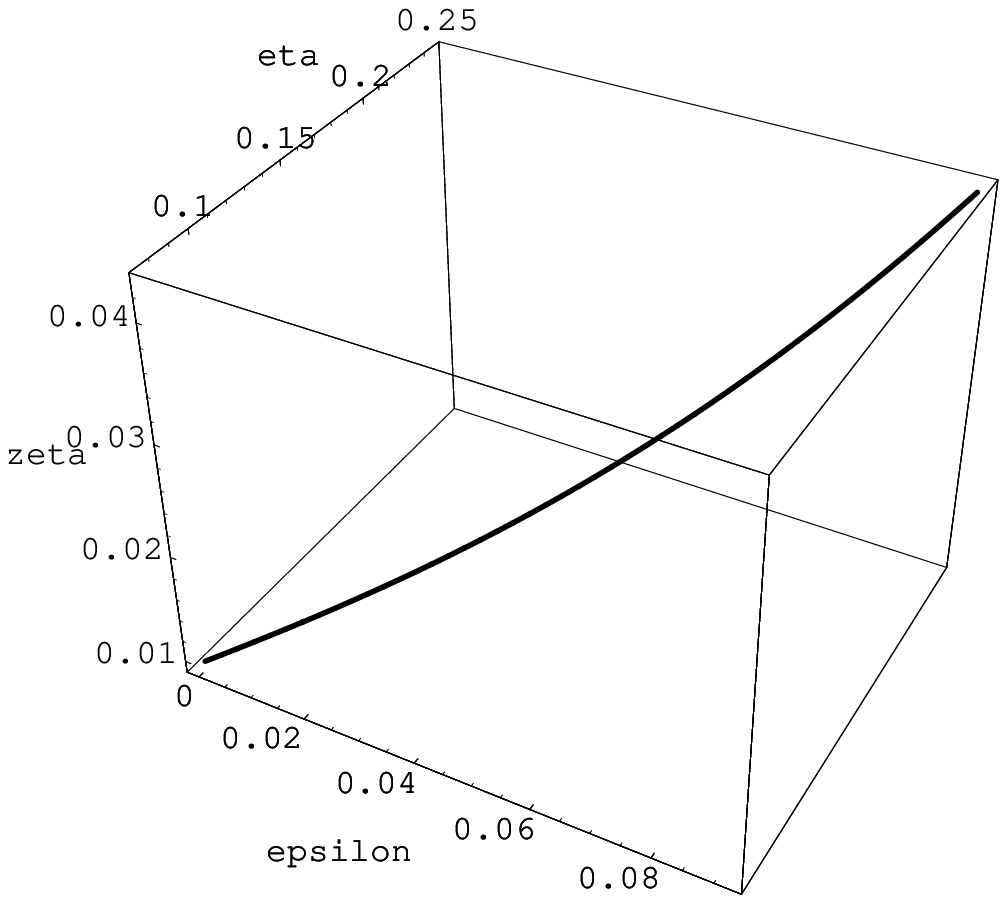}
\epsfxsize=0.3\columnwidth \epsfbox{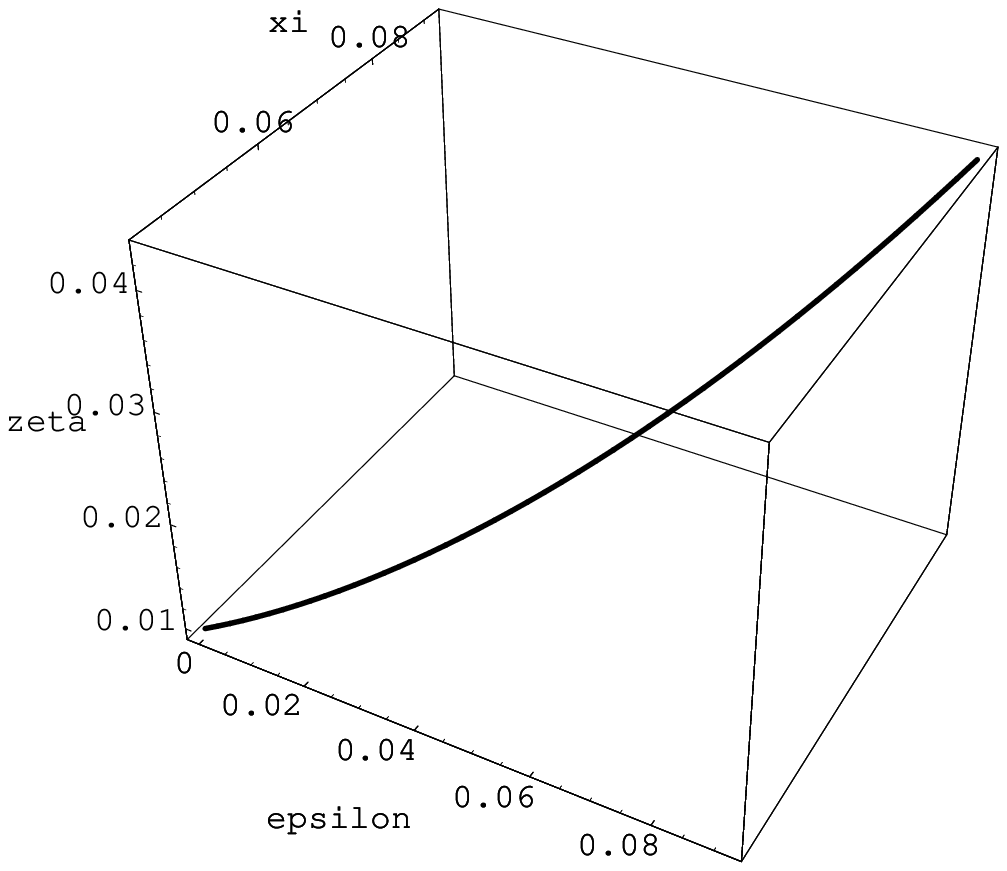}
\end{center}
\caption{Ignoring $\theta$ only, constraints on the slow-roll
parameters are showed for $n_s=1.21$, $\alpha_s=-0.102$ and
$\beta_s=0.0318$, where $r\simeq 16\epsilon\leq 1.5$.}
\end{figure}

A premature truncation of the slow-roll hierarchy appears if
$\zeta\sim {\cal O}(10^{-3})$. For example, for $\zeta<10^{-2}$,
constraints on the other slow-roll parameters are respectively
$\epsilon\in[0,0.002]$, $\eta\in[0.075,0.078]$ and $\xi\simeq
0.042$. Now the tensor-scalar ratio is smaller than $0.04$.

We also investigate the evolution of the slow-roll parameters and
the spectral index and its running. The similar behaviors of these
parameters are obtained for taking different values of slow-roll
parameters at $k=0.002$Mpc$^{-1}$ along the line in fig. 3. For
example, we take $\epsilon=0.02$, $\eta=0.112$, $\xi=0.047$ and
$\zeta=0.014$ and the evolution of the parameters are showed in fig.
4.
\begin{figure}[h]
\begin{center}
\epsfxsize=0.6\columnwidth \epsfbox{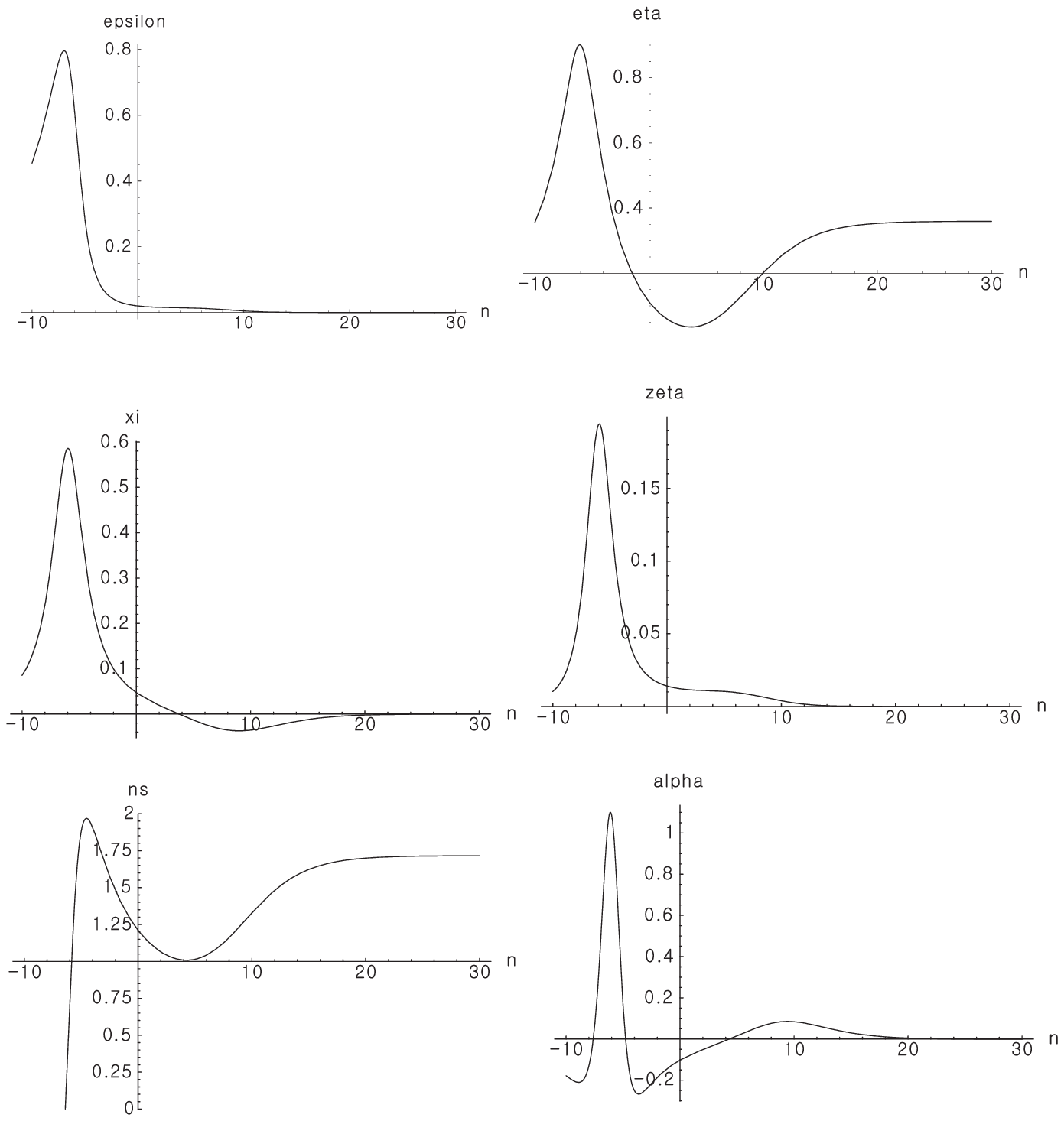}
\end{center}
\caption{Ignoring $\theta$ only, the evolution of the slow-roll
parameters and spectral index and its running are showed.}
\end{figure}
An interesting result is that all of the slow-roll parameters go to
zero within roughly 15 e-folds after $k=0.002$Mpc$^{-1}$ leaves the
horizon, except $\eta$ which approaches to a positive constant
$\eta_c$. Since $\eta_c$ is still much less than 3, the slow-roll
inflation will not end. How to end inflation will be discussed in
section 4.

We need to make one point clear here. If we only consider the
spectral index and the running to reconstruct the slow-roll
parameters, the constraints is looser, since there are four
parameters, but only two constraints. Taking the fourth slow-roll
parameter for a nice fit to the running of running, we find that a
large enough number of e-folds is obtained.

\subsection{$Summary$}

To summarize, the slow-roll reconstruction shows that one more
parameter is needed if we want to fit the observational parameters,
e.g. three slow-roll parameters is needed to fit the spectral index
and its running and four slow-roll parameters are required if we
also want to fit the running of running. Including the fourth
slow-roll parameter $\zeta$, a tentative truncation of the slow-roll
hierarchy is obtained for a small value of $\epsilon$; otherwise, we
should take care of the contribution from higher order slow-roll
parameters when performing a fit to the spectral index and its
running. We also need to keep in mind that we still have not built
such an inflationary model from a fundamental theory.

\setcounter{equation}{0}
\section{Spectral index in noncommutative inflation model}

There is not a realistic inflationary model providing a negative
running of the spectral index with large enough absolute value. One
question is what is the origin of the running spectral index with
the viewpoint of fundamental theories. In
\cite{hg,hld,bhm,hla,hlb,hlc,pncf}, the authors proposed that the
inflationary models in noncommutative spacetime open a window to
improve the fit to the running spectral index for the typical
inflationary models. The other relevant trans-Planckian physics
\cite{prunt} and non-minimal coupled inflation model \cite{ml} have
also been discussed widely.

In this section, we compute the amplitude of primordial scalar and
tensor power spectra in noncommutative inflation model and discuss
the truncation of the slow-roll hierarchy in order to fit the
running spectral index.

\subsection{Spectral index in noncommutative inflation model}

Noncommutative spacetime naturally emerges in string theory
\cite{ncst}, which implies a new uncertainty relation \e
\label{urst}\Delta t_p \Delta x_p \geq l_s^2, \q where $t_p$ and
$x_p$ are the physical time and space, $l_s$ is the uncertainty
length scale, or string scale in string theory. To make this paper
self-consistent, we briefly review the calculation of the primordial
power spectrum \cite{hlc} for noncommutative inflationary models.
The spacetime noncommutative effects are encoded in a new product
among functions, namely the star product, replacing the usual
algebra product. Since the evolution of the background and inflaton
is homogeneous, the standard classical cosmological equations does
not change.

To make the uncertainty relation in (\ref{urst}) clearer in FRW
background, we introduce a new time coordinate $\tau$ as \e
ds^2=dt^2-a^2(t)d{\vec{x}}^2=a^{-2}(\tau)d\tau^2-
a^2(\tau)d{\vec{x}}^2.\q Now the uncertainty relationship
(\ref{urst}) for the coordinates in the above metric becomes \e
\label{ustc}\Delta \tau \Delta x \geq l_s^2. \q The star product can
be explicitly defined as \e
\label{dstp}f(\tau,x)*g(\tau,x)=e^{-{i\over 2}l_s^2(\p_x \p_{\tau'}
- \p_{\tau} \p_{x'})}f(\tau, x)g(\tau', x')|_{\tau'=\tau, x'=x}.\q
Because the comoving curvature perturbation $\cal R$ not only
depends on time, but also depends on the position in the space, the
equation of motion for $\cal R$ is modified by the noncommutative
effects, \e \label{emsp}u_k''+\(k^2-{z_k'' \over z_k} \)u_k=0, \q
where \m \label{scaf} z_k^2(\te)&=&z^2y_k^2(\te), \quad
y_k^2=(\beta_k^+\beta_k^-)^{\half},\\ {d\te \over
d\tau}&=&\left({\beta_k^-\over \beta_k^+}\right)^\half,\quad
\beta_k^\pm =\half (a^{\pm 2}(\tau+\l_s^2k)+a^{\pm
2}(\tau-l_s^2k)),\nonumber \n $z=a\dot \phi/H$, ${\cal R}_k(\te) =
u_k(\te) / z_k(\te)$ is the Fourier modes of $\cal R$ in momentum
space and the prime in (\ref{emsp}) denotes the derivative with
respect to the modified conformal time $\te$. The deviation from the
commutative case encodes in $\beta^\pm_k$ and the corrections from
the noncommutative effects can be parameterized by ${Hk\over
aM_s^2}$. After a lengthy but straightforward calculation, we get \m
\label{zkt}{z_k'' \over
z_k}&=&2(aH)^2\(1+\epsilon-{3\over 2}\eta-2\mu+{\cal O}(\xi,\epsilon^2,\epsilon\eta,\eta^2) \), \\
aH&\simeq& {-1\over \te}(1+\epsilon+\mu),\nonumber \n where \e
\label{du}\mu = {H^2 k^2 \over a^2 M^4_s}\q is the noncommutative
parameter and $M_s = l_s^{-1}$ is the noncommutative mass scale or
string mass scale. Solving eq. (\ref{emsp}) yields the amplitude of
the scalar comoving curvature fluctuations in noncommutative
spacetime \e \label{asc}\Delta_{\cal R}^2={k^3\over
2\pi^2}\left|{\cal R}_k(\te)
\right|^2=\(1-(2C+1)\epsilon+C\eta\)^2{H^2/M_p^2 \over
8\pi^2\epsilon} (1+\mu)^{-4}, \q where $H$ takes the value when the
fluctuation mode $k$ crosses the Hubble radius ($z_k''/z_k=k^2$) and
$k$ is the comoving Fourier mode. Using (\ref{du}) and (\ref{dkp}),
we obtain \e {d\mu \over d\ln k}={-4\epsilon\mu\over 1-\epsilon}.\q
The spectral index and its running and running of running are
respectively
\m n_s&=&1-4\epsilon+2\eta-2C\xi-8(C+1)\epsilon^2+(6+10C)\epsilon\eta\nonumber \\ &+&16(\epsilon+\epsilon^2)\mu,\label{nns}\\
\alpha_s&=&-2\xi-8\epsilon^2+10\epsilon\eta-(8+14C)\epsilon\xi+2C\eta\xi+2C\zeta\nonumber \\ &-&32(\epsilon^2+\epsilon\eta)\mu,\label{nrun}\\
\beta_s&=&-14\epsilon\xi+2\eta\xi-2C\xi^2+2\zeta-2C\theta\nonumber
\\
&-&32\epsilon^3+62\epsilon^2\eta-20\epsilon\eta^2-(46+56C)\epsilon^2\xi+(26+48C)\epsilon\eta\xi
\nonumber \\
&-&2C\eta^2\xi+(10+20C)\epsilon\zeta-6C\eta\zeta\nonumber
\\ &+&160\epsilon\eta^2\mu+64\epsilon^2\eta\mu+32\epsilon\xi\mu. \label{nrrun}\n
Similarly, the amplitude of the tensor perturbations and the
tensor-scalar ratio are respectively given by \m
\Delta_T^2&=&\(1-(C+1)\epsilon \)^2{H^2/M_p^2\over
\pi^2/2}(1+\mu)^{-4}, \\ r&=&{\Delta_T^2\over \Delta_{\cal
R}^2}=16\epsilon(1+2C(\epsilon-\eta)). \n When $\mu=0$ or $H\ll
M_s$, all of the above results in noncommutative inflation are just
the same as those in the standard slow-roll inflation model.

\subsection{Slow-roll reconstruction and slow-roll hierarchy in noncommutative inflation model}

In this subsection, we reconstruct the slow-roll parameters and the
noncommutative parameter $\mu$ by fitting the running spectral index
from WMAP three-year data.

First, we consider the case with $\xi=\zeta=\theta=0$. Now there are
three free parameters: $\epsilon$, $\eta$ and $\mu$. Fitting the
spectral index and its running and running of running, we find
$\epsilon=0.077$, $\eta=0.072$, $\mu=0.3$ and the tensor-scalar
ratio is 1.23. At first sight, this result seems trivial, because we
use three free parameters to fit three observed parameters. But the
non-trivial thing is that the slow-roll parameters and $\mu$ are
evaluated within the reasonable range. The evolution of these
parameters are figured out in fig. 5.
\begin{figure}[h]
\begin{center}
\epsfxsize=0.6\columnwidth \epsfbox{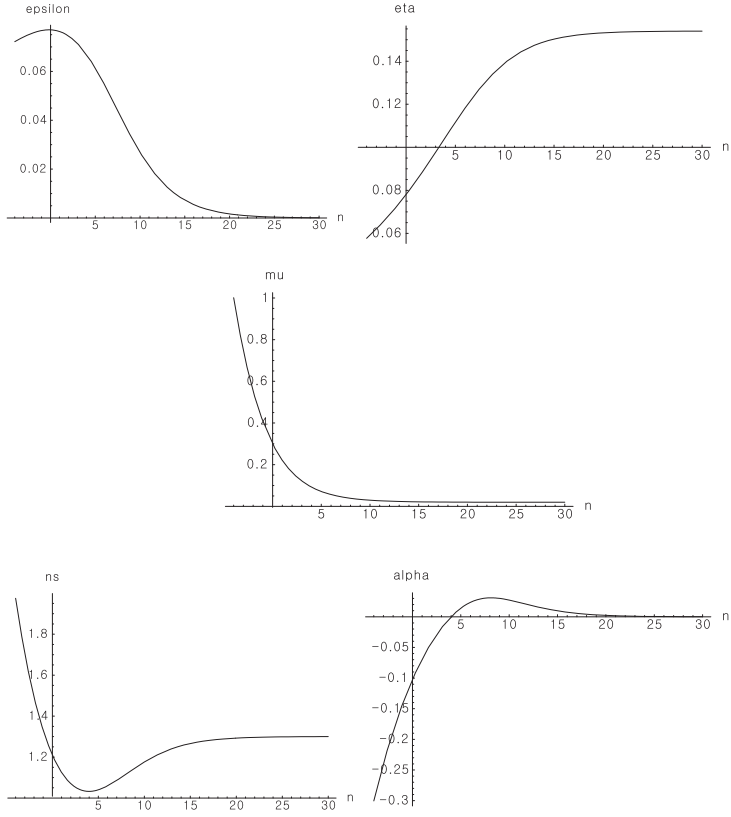}
\end{center}
\caption{Setting $\xi=\zeta=\theta=0$, the evolution of the
slow-roll parameters and spectral index and its running are showed.}
\end{figure}
After fluctuation mode $k=0.002$Mpc$^{-1}$ stretched outside the
Hubble horizon, both $\epsilon$ and $\mu$ approach zero; however,
$\eta$ goes to a positive constant $\eta_c$ which is much smaller
than 3. The mechanism to end inflation will be discussed in section
4.

Second, we consider the case where only $\zeta$ and $\theta$ are set
to zero. The constraints on the slow-roll parameters and $\mu$ to
fit the spectral index, its running and running of running are shown
in fig. 6.
\begin{figure}[h]
\begin{center}
\leavevmode
\epsfxsize=0.3\columnwidth \epsfbox{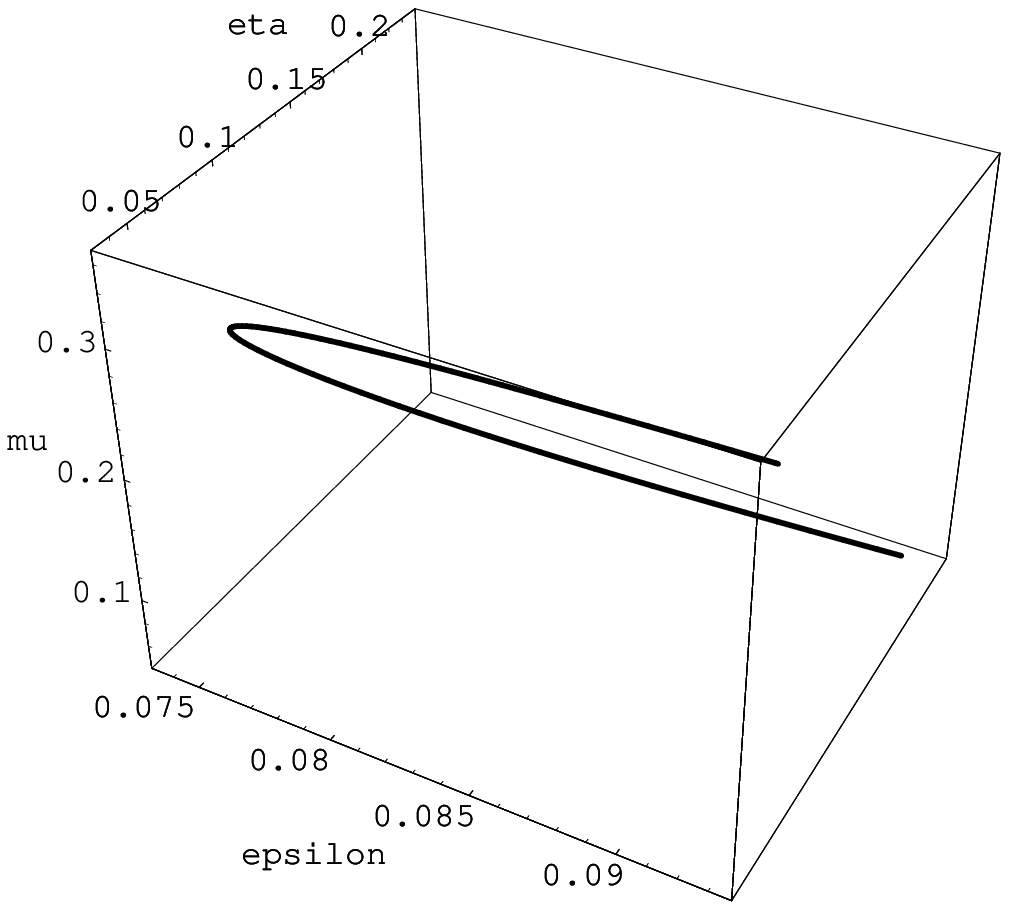}
\epsfxsize=0.3\columnwidth \epsfbox{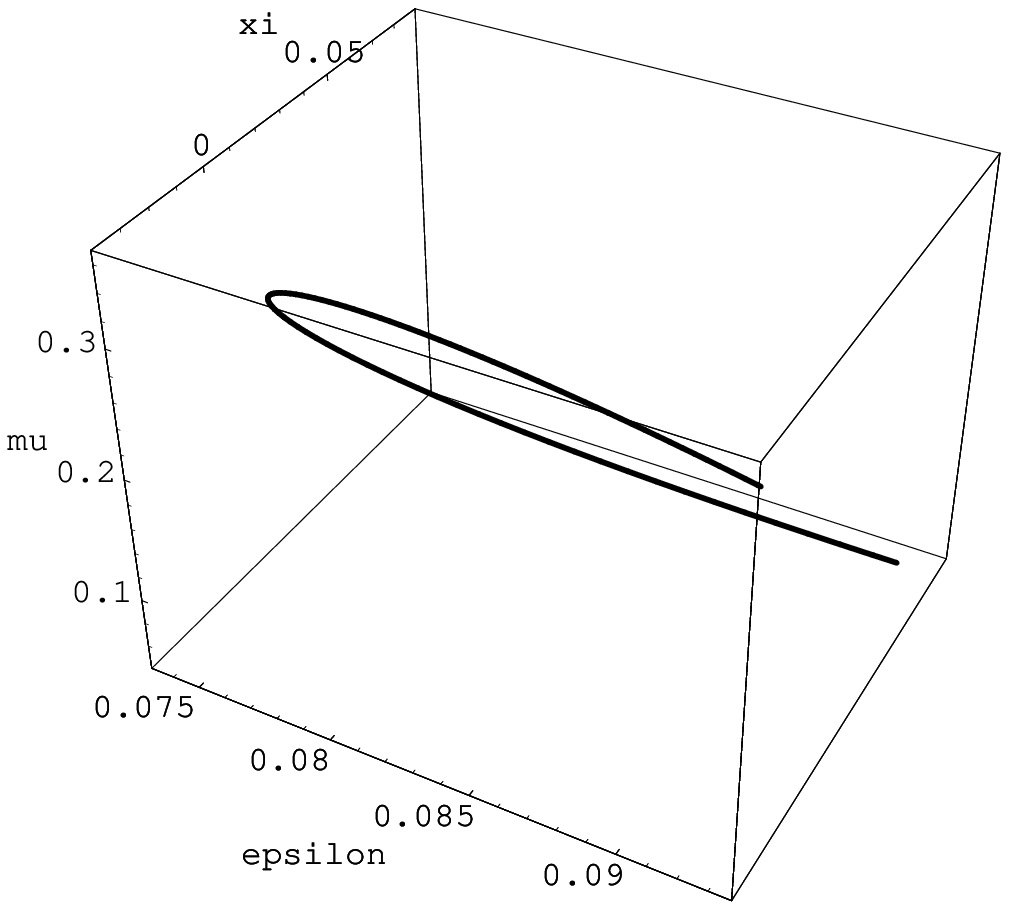}
\end{center}
\caption{Ignoring $\zeta$ and $\theta$, the constraints on the
slow-parameters and $\mu$ are showed fitting to $n_s=1.21$,
$\alpha_s=-0.102$ and $\beta_s=0.0318$.}
\end{figure}
We are also interested in the evolution of the slow-roll parameters
and spectral index and its running. We scan the corresponding value
of the slow-roll parameters and $\mu$ in fig. 6. If we require that
slow-roll inflation lasts more than 20 e-folds after CMB scales
leave the horizon, the constraints on these parameters are
$\epsilon\in[0.076,0.094]$, $\eta\in[0.044,0.078]$,
$\xi\in[-0.039,0.005]$ and $\mu\in[0.284,0.352]$ at
$k=0.002$Mpc$^{-1}$. Within the permitted region for these
parameters at $k=0.002$Mpc$^{-1}$, for example, we take
$\epsilon=0.081$, $\eta=0.062$, $\xi=-0.01$ and $\mu=0.318$, and the
evolution of the slow-roll parameters and spectral index and its
running are shown in fig. 7.
\begin{figure}[h]
\begin{center}
\epsfxsize=0.6\columnwidth \epsfbox{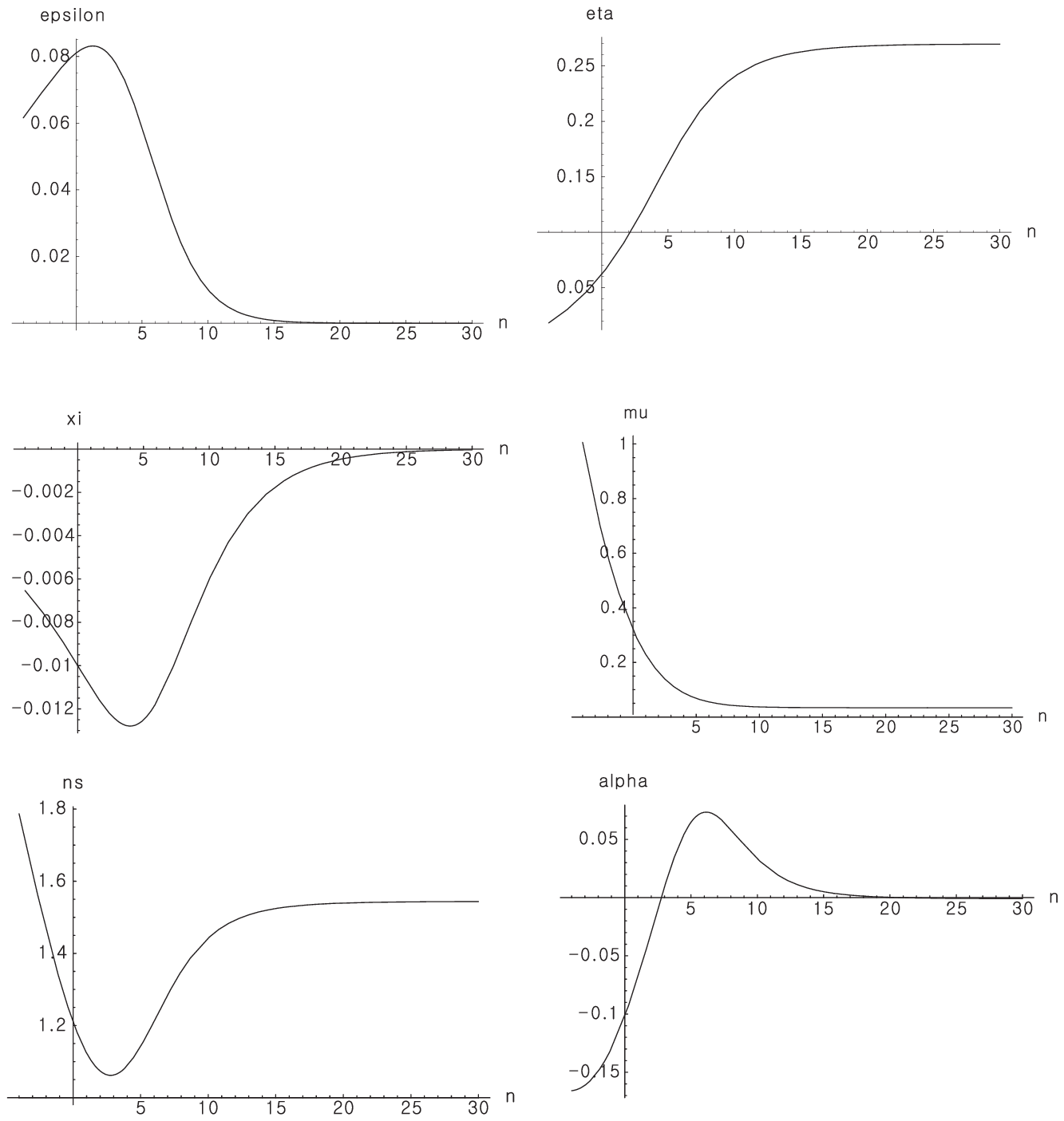}
\end{center}
\caption{Setting $\zeta=\theta=0$, the evolution of the slow-roll
parameters and spectral index and its running are showed.}
\end{figure}
Similar behavior for $\epsilon$, $\eta$, $\xi$ and $\mu$ is obtained
for the other points along the lines in fig. 6. As a universal
result, the slow-roll parameters go to zero, except $\eta$ which
approaches a positive value $\eta_c$.

To summarize, noncommutative inflation can provide a running
spectral to fit WMAP three-year data and the slow-roll hierarchy is
obtained. A feature for the noncommutative inflation model is that a
large amplitude of the tensor perturbations is required ($r\geq
1.22$).

\setcounter{equation}{0}
\section{The last stage of inflation}

In section 2 and 3, requiring slow-roll inflation lasts more than
twenty e-folds after CMB scales leave the horizon, we reconstructed
the slow-roll parameters for standard inflation model and
noncommutative inflation respectively. Our numerical results show
that the slow-roll parameters approach zero in the last stage of
inflation, except $\eta$ which goes to a positive constant $\eta_c$.
Assume that the value of inflaton and Hubble parameter are
respectively $\eta_c$ and $H_c$ at the end of inflation, eq.
(\ref{hht}) can be written as \e \eta_c=2M_p^2{1\over H}{d^2H\over
d\phi^2}. \label{sht}\q Since Hubble parameter $H$ and the value of
$\phi$ monotonically decrease, eq. (\ref{sht}) becomes \e {dH\over
d\phi}=\({\eta_c\over 2M_p^2}(H^2-H_c^2)\)^{1/2}.\q The solution of
above equation is given by \e H=H_c\cosh\(\sqrt{\eta_c\over
2}{\phi-\phi_c\over M_p}\)\quad \hbox{for}\quad\phi>\phi_c.\q Since
$\epsilon\rightarrow 0$ and then $\dot \phi\rightarrow 0$ in the
last stage of inflation, the asymptotical behavior of the potential
$V(\phi)$ takes the form \e V(\phi)\rightarrow
3M_p^2H^2=V_0\cosh^2\(\sqrt{\eta_c\over 2}{\phi-\phi_c\over
M_p}\),\q where $V_0=3M_p^2H_c^2$. This potential can be trusted
only when $\phi\rightarrow \phi_c$ and then we expand this potential
around $\phi_c$ as \e V(\phi)\simeq V_0\(1+{\eta_c\over
2}{(\phi-\phi_c)^2\over M_p^2}\). \label{lpot}\q We need to stress
that this is only the potential dominated the last stage of
inflation. We can not figure out a simple function of the potential
for the stage with running spectral index.

Since $\eta_c\ll 3$, inflation does not end by violating the
slow-roll conditions. As we know, hybrid inflation provides a
natural mechanism to end this kind of inflation \cite{hybrid}. A
tentative potential for hybrid inflation can be \e
V(\phi,\varphi)=V_0\(1+{\eta_c\over 2}{(\phi-\phi_c)^2\over
M_p^2}\)+\half g(\phi^2-\phi_c^2)\varphi^2+{\lambda\over
4}\varphi^4, \q where both $g$ and $\lambda$ are positive constant.
During the period of inflation, $\phi>\phi_c$ and $\varphi$ stays at
its global minimum $\varphi=0$. When inflaton $\phi$ goes to
$\phi_c$, the end of inflation is triggered by $\varphi$.

\setcounter{equation}{0}
\section{Conclusion}

In this paper, the slow-roll parameters are reconstructed in the
standard inflationary models and the noncommutative inflationary
models. Requiring that slow-roll inflation lasts more than 20
e-folds after CMB scales leave the horizon, we get a stringent
constraint on the slow-roll parameters in order to fit the running
spectral index: the fourth slow-roll parameter is needed for usual
inflationary models, or a large tensor-scalar ratio with $r\geq
1.22$ is needed for noncommutative inflationary models. Thus if
$r\ll 1$, the standard inflationary model is favored; but the
noncommutative seems better in the case with $r\geq 1$. In both
cases, the potential of the inflaton in the last stage of inflation
should take the form of eq. (\ref{lpot}) and hybrid type inflation
provides a natural mechanism to end the inflation. Combining with
SDSS, the tensor-scaler ratio should be less than $0.67$. If so, the
noncommutative inflationary model can not help us to get a negative
running of the spectral index with large absolute value.

However the running spectral is still not strongly favored by
observations. We hope Planck or further data from WMAP can tell us
whether the spectral index does run or not. If it really runs, the
observations can tell us more details about inflation.

\vspace{.5cm}

\noindent {\bf Acknowledgments}

We would like to thank S. Kim, E.D. Stewart for useful discussions.

\end{document}